\RequirePackage{fix-cm}
\documentclass[smallextended]{svjour3}       
\smartqed  
\usepackage{graphicx}
\journalname{Eur. Phys. J. C}
\usepackage{bm}
\usepackage{slashed}
\usepackage{url}
\usepackage{float}

\usepackage{amsmath}
\usepackage{comment}
\usepackage{subfigure}
\usepackage{float}

\begin{document}


\title{Confinement in a three-dimensional Yang-Mills theory}



\author{Marco Frasca         
}


\institute{Marco Frasca \at
              Via Erasmo Gattamelata, 3 \\
							00176 Rome (Italy)
              \email{marcofrasca@mclink.it}           
}

\date{Received: date / Accepted: date}

\maketitle

\begin{abstract}
We show that, starting from known exact classical solutions of the Yang-Mills theory in three dimensions, the string tension is obtained and the potential is consistent with a
marginally
confining theory. The potential we obtain agrees fairly well with preceding findings in literature but here we derive it analytically from the theory without further assumptions. The string tension is in strict agreement with lattice results and the well-known theoretical result by Karabali-Kim-Nair analysis. Classical solutions depend on a dimensionless numerical factor arising from integration. This factor enters into the determination of the spectrum and has been arbitrarily introduced in some theoretical models. We derive it directly from the solutions of the theory
and is now fully justified.
%
The agreement obtained with the lattice results for the ground state of the theory is well below 1\% at any value of the degree of the group. 
\end{abstract}


\section{Introduction}

A deep understanding of Yang-Mills theory in all the range of the coupling represents a fundamental aspect of our comprehension of strong interactions. The reason is that this would open the possibility to accomplish computations of the behavior of the theory in the low-energy limit where the theory displays bound states. Currently, the only way to obtain results that are derived directly from the theory is through extensive use of lattice computations on large computer facilities. This has permitted for the Yang-Mills theory to obtain both the spectrum and the behavior of propagators in several gauges also at finite temperature \cite{Durr:2008zz,Bazavov:2009bb,Maas:2011se,Petreczky:2012rq} in four dimensions and similarly for the case $d=2+1$ \cite{Teper:1998te,Lucini:2002wg,Bringoltz:2006zg,Caselle:2011fy,Caselle:2011mn,Athenodorou:2016ebg}. Specially in this latter case, very precise results exist for the string tension and the spectrum.

From a theoretical standpoint, the situation appears decisively better for the three-dimensional case where some analysis have been performed producing excellent agreement with lattice computations for the string tension \cite{Karabali:1997wk,Karabali:1998yq,Nair:2002yg,Karabali:2009rg} and the spectrum \cite{Leigh:2005dg,Leigh:2006vg}. The starting point was a work by Karabali, Kim and Nair that proposed a proper set of matrix variables to work with in this case to put forward a wavefunction and derive fundamental results of the theory \cite{Karabali:1995ps,Karabali:1997wk,Karabali:1998yq}. Karabali, Kim and Nair approach appears greatly successful in the derivation of the string tension and higher order corrections \cite{Karabali:2009rg}. For the spectrum, a different wavefunction was postulated \cite{Leigh:2005dg,Leigh:2006vg} always in the framework of Karabali, Kim and Nair formalism. Again, the agreement with lattice data was impressive.

In a recent paper of ours we were able to build a quantum field theory for the self-interacting scalar field in the limit of the coupling running to infinity \cite{Frasca:2013tma}. We displayed a set of classical solutions that, notwithstanding we started from a massless equation, showed a massive dispersion relation. These solutions were already proposed in \cite{Frasca:2009bc} but the idea in \cite{Frasca:2013tma} was to consider them as the vacuum expectation value of the field and build the quantum theory from them. In this way one has that conformal invariance is broken and a zero mode appears. The particles get a mass and a tower of excited states described by the spectrum of a harmonic oscillator. This theory shares a trivial infrared fixed point and an ultraviolet trivial fixed point making the theory overall trivial but with a mass gap. On this ground it is a natural question to ask if also a Yang-Mills theory can share such classical solutions and a corresponding quantum field theory built upon them. The answer was affirmative as we showed in \cite{Frasca:2009yp} but this is true asymptotically in the general case while the result holds exactly just in the Lorenz (Landau) gauge. The corresponding quantum field theory develops a mass gap but is trivial at both sides of the range due to the trivial infrared fixed point and asymptotic freedom on the other side. This scenario has received some confirmations recently in a work by Deur \cite{Deur:2016bwq}. The idea is to use the propagator of the scalar theory and compute the potential with a running coupling as expected in a Yang-Mills theory. The agreement with a confining potential obtained from lattice computations is striking. The running coupling in Yang-Mills theories has been widely discussed in \cite{Nesterenko:1999np,Nesterenko:2001st,Nesterenko:2003xb,Baldicchi:2007ic,Baldicchi:2007zn,Bogolubsky:2009dc,Duarte:2016iko,Deur:2016tte}.

In this paper we develop this approach, alternative to the Nair and Karabali formalism, deriving all the properties of the theory. That is, we solve the Yang-Mills theory in 2+1 dimensions in another way and we will get results in
strikingly good
agreement with lattice data, validating this approach. As a by-product we will get an important hint that the flux tube description of hadron emerging in AdS/CFT approach \cite{Brodsky:2014yha} is a successful one as we are able to get the right ground state of the theory by introducing the same factor as in the Isgur-Paton theory \cite{Isgur:1984bm} as demonstrated by Teper and Johnson \cite{Johnson:2000qz}. Anyhow,
it is important to point out that we show how the correction factor, arbitrarily introduced in the aforementioned works, is here properly obtained by solving the classical equations of motions, being this just an integration constant. This kind of arbitrariness enters into Yang-Mills theory and the scalar field theory, to which it maps, due to the properties of the differential equations of the theories. Another result that is really interesting with our approach is that our confining potential is almost identical to the one obtained in \cite{Leigh:2005dg,Leigh:2006vg}. These authors obtained it after some hypotheses to be verified and we show here that they were correct.
The three-dimensional theory is so proved to be marginally confining.
%

The value of this analysis can be just of mathematical interest to study the structure of a quantum field theory in lower dimensions and to get some understanding of the behavior of the four dimensional theory starting from a simpler case. Indeed, as stated above, a lot of effort has been spent, both through lattice and theoretical study, to analyze the three dimensional case. Indeed, some of the mathematical techniques devised for this case could be successfully applied to the more realistic four dimensional case.

The paper is so structured. In Sec.~\ref{sec1} we introduce a set of exact classical solutions to work with for the quantum field theory. In Sec.~\ref{sec2} we derive the gluon propagator for the classical theory.
In Sec.~\ref{sec22} we discuss the ghost sector of the theory.
In Sec.~\ref{sec3} we evaluate the quantum corrections. In Sec.~\ref{sec4} we compute the Wilson loop obtaining a confining potential in agreement with literature and the string tension in agreement with lattice data. In Sec.~\ref{sec5} we comnpute the ground state of the theory giving the lowest glueball state. Finally, in Sec.~\ref{sec6} the conclusions are presented.

\section {Classical solutions \label{sec1}
}


Motion equation for Yang-Mills theory can be straightforwardly written down for any number of dimensions and SU(N) group in the form
\cite{Chaichian:1984mm}
\begin{equation}
\label{eq:YM}
    {\cal D}^\mu F_{\mu\nu}=0
\end{equation}
being
\begin{equation}
   {\cal D}_\mu=\partial_\mu-igT^aA^a_\mu
\end{equation}
the covariant derivative, $T_a$ the generators of the group and $A^a_\mu$ the potentials ($a,b,c,\ldots$ are color indexes running from 1 to
$N^2-1$
), and
\begin{equation}
    F_{\mu \nu}^a = \partial_\mu A_\nu^a-\partial_\nu A_\mu^a+gf^{abc}A_\mu^bA_\nu^c
\end{equation}
the field components with $F_{\mu\nu}=T^aF^a_{\mu\nu}$ and $f^{abc}$ the structure constants of the group. As our aim is to work out a result in quantum field theory, we add a term into eq.(\ref{eq:YM}) to fix the gauge in the form
\begin{equation}
   -\left(1-\frac{1}{\xi}\right)\partial_\nu(\partial\cdot A^a)
\end{equation}
with $\xi$ a free parameter determining the gauge choice.

Using perturbation theory, one can show that there exists a set of solutions of Yang-Mills equations of motion that can be cast in the form \cite{Frasca:2009yp}
\begin{equation}
\label{eq:asymp}
   A_\mu^a(x)=\eta_\mu^a\chi(x)+O\left(1/Ng^2\right).
\end{equation}
being $\eta_\mu^a$ a set of constants to be determined depending on the problem at hand
(e.g., for SU(2) in the Landau gauge, one can take $\eta_1^1=\eta_2^2=\eta_3^3=1$, all other components being zero).
%
Putting these potentials into the equations of motion yields \cite{Frasca:2009yp}
\begin{equation}
\label{eq:scal}
   \partial^2\chi(x)+Ng^2\chi^3(x)=0.
\end{equation}
%
These solutions become exact and not just perturbative for the Lorenz (Landau) gauge. An interesting aspect of these solutions is that hold in any dimensions $d>2$. For $d=2$ Yang-Mills equations of motion are trivial and no such solutions can be found.

Without exploiting all the possible solutions of eq.(\ref{eq:scal}) we limit our interest to a subclass of solutions that have the property to be massive even if we started from massless equations of motion. We have fully exploited this case in Ref.\cite{Frasca:2013tma}. In this paper we consider such exact solutions as a ground state of the quantum field theory of a scalar field. In 3+1 dimensions this can be written down as \cite{Frasca:2013tma}
\begin{equation}
    \chi_{d=3+1}(x)=\mu\left(\frac{2}{Ng^2}\right)^\frac{1}{4}\operatorname{sn}\left(k\cdot x+\phi,-1\right)
\end{equation}
being sn a Jacobi elliptic function, $\phi$ an arbitrary phase, $\mu$ an arbitrary constant having the dimension of a mass and provided that
\begin{equation}
    k^2=\sqrt{\frac{Ng^2}{2}}\mu^2.
\end{equation}
So, if we interpret $k$ as a four-vector of momenta, this can be seen as the dispersion relation of a massive wave. These solutions are rather counterintuitive as we started from a pure massless theory. A mass term can be seen to arise from the nonlinearities of the equations we started from. In the following we will assume that such solutions are just the ground state for the quantum field theory we aim to study.

%


In 2+1 dimensions Yang-Mills equations have a coupling $g^2$ having the dimension of a mass or inverse of a length. This means that our solution takes the form
\begin{equation}
\label{eq:exsol}
		\chi_{d=2+1}(x)=a\cdot 2^\frac{1}{4}\sqrt{Ng^2}\operatorname{sn}\left(k\cdot x+\phi,-1\right)
\end{equation}
being $a$ an arbitrary dimensionless constant to be fixed in the quantum theory and $\phi$ an arbitrary phase. This holds provided the following dispersion relation holds
\begin{equation}
\label{eq:ds2+1}
    k^2=a^2\cdot\frac{N^2g^4}{\sqrt{2}}.
\end{equation}

%

\section{Gluon propagator \label{sec2}}

We need to introduce the propagator of Yang-Mills theory in the infrared limit. This is generally accomplished by a current expansion \cite{Frasca:2013tma,Cahill:1985mh}. Instead to start from the action, we prefer the equations of motion \cite{Rubakov:2002fi}
\begin{eqnarray}
&&\partial^\mu\partial_\mu A^a_\nu-\left(1-\frac{1}{\alpha}\right)\partial_\nu(\partial^\mu A^a_\mu)+gf^{abc}A^{b\mu}(\partial_\mu A^c_\nu-\partial_\nu A^c_\mu)+gf^{abc}\partial^\mu(A^b_\mu A^c_\nu) \nonumber \\ 
&&+g^2f^{abc}f^{cde}A^{b\mu}A^d_\mu A^e_\nu = j^a_\nu.
\end{eqnarray}
Then, we assume a functional form $A^a_\nu=A^a_\nu[j]$ and perform a Taylor expansion around the asymptotic solution (\ref{eq:asymp}). We have to take in mind that, for the Landau gauge, these solutions are exact but just asymptotic for whatever other gauge choice. So, we take in general
\begin{equation}
   A^a_\nu[j(x)]=\eta_\nu^a\chi(x)+\int d^dx'\left.\frac{\delta A_\nu^a}{\delta j_\mu^b(x')}\right|_{j=0}j_\mu^b(x')+
	\frac{1}{2}\int d^dx'd^dx''\left.\frac{\delta^2 A_\nu^a}{\delta j_\mu^b(x')\delta j_\kappa^c(x'')}\right|_{j=0}
	j_\mu^b(x')j_\kappa^c(x'')+\ldots.
\end{equation}
We are assuming here that eq.(\ref{eq:asymp}) represents the ground state of the theory i.e. $A^a_\nu[0]=\eta_\nu^a\chi(x)$. These describe oscillations around a vacuum expectation value of the fields as seen from our solutions. Then, the propagator of the theory will be
\begin{equation}
   G_{\mu\nu}^{ab}(x,x')=\left.\frac{\delta A_\nu^a(x)}{\delta j_\mu^b(x')}\right|_{j=0}.
\end{equation}
We can obtain the corresponding equation by doing the functional derivative on the equation of motion. We get
\begin{eqnarray}
&&\partial^2 \frac{\delta A_\nu^a(x)}{\delta j_\rho^e(x')}
-\left(1-\frac{1}{\alpha}\right)\partial_\nu\left(\partial^\mu\frac{\delta A_\mu^a(x)}{\delta j_\rho^e(x')}\right) \nonumber \\
&&+gf^{abc}\frac{\delta A_\mu^b(x)}{\delta j_\rho^e(x')}\left(\partial^\mu A^c_\nu-\partial_\nu A^{\mu c}\right) \nonumber \\
&&+gf^{abc}A_\mu^b\left(\partial^\mu\frac{\delta A^c_\nu(x)}{\delta j_\rho^e(x')}
-\partial_\nu\frac{\delta A^{\mu c}(x)}{\delta j_\rho^e(x')}\right) \nonumber \\
&&+gf^{abc}\partial^\mu\left(\frac{\delta A_\mu^b(x)}{\delta j_\rho^e(x')} A^c_\nu\right) 
+gf^{abc}\partial^\mu\left(A_\mu^b\frac{\delta A_\nu^c(x)}{\delta j_\rho^e(x')}\right) \\ \nonumber
&&+g^2f^{abc}f^{cdh}\frac{\delta A_\mu^b(x)}{\delta j_\rho^e(x')}A^d_\mu A^h_\nu \\ \nonumber
&&+g^2f^{abc}f^{cdh}A^{b\mu}\frac{\delta A_\mu^d(x)}{\delta j_\rho^e(x')} A^h_\nu \\ \nonumber
&&+g^2f^{abc}f^{cdh}A^{b\mu}A^d_\mu\frac{\delta A_\nu^h(x)}{\delta j_\rho^e(x')}
= \delta_{ae}\eta_{\nu\rho}\delta^d(x-x').
\end{eqnarray}
Imposing $j=0$ one obtains the following equation for the Green function of Yang-Mills theory
\begin{eqnarray}
&&\partial^2G_{\nu\rho}^{ae}(x,x')
-\left(1-\frac{1}{\alpha}\right)\partial_\nu\partial^\mu G_{\mu\rho}^{ae}(x,x') \nonumber \\
&&+gf^{abc}G_{\mu\rho}^{be}(x,x')\left(\partial^\mu A^c_\nu-\partial_\nu A^{\mu c}(x)\right) \nonumber \\
&&+gf^{abc}A_\mu^b\left(\partial^\mu G_{\nu\rho}^{ce}(x,x')
-\partial_\nu G_{\mu\rho}^{ce}(x,x')\right) \nonumber \\
&&+gf^{abc}\partial^\mu \left(A^c_\nu G_{\mu\rho}^{be}(x,x')\right)
+gf^{abc}\partial^\mu\left(A_\mu^b G_{\nu\rho}^{ce}(x,x')\right) \\ \nonumber
&&+g^2f^{abc}f^{cdh}G_{\mu\rho}^{be}(x,x')A^{\mu d} A^h_\nu \\ \nonumber
&&+g^2f^{abc}f^{cdh}A^{b\mu}G_{\mu\rho}^{de}(x,x') A^h_\nu \\ \nonumber
&&+g^2f^{abc}f^{cdh}A^{b\mu}A^d_\mu G_{\nu\rho}^{he}(x,x')
= \delta_{ae}\eta_{\nu\rho}\delta^d(x-x').
\end{eqnarray}
or
\begin{eqnarray}
&&\partial^2G_{\nu\rho}^{ae}(x,x')
-\left(1-\frac{1}{\alpha}\right)\partial_\nu\partial^\mu G_{\mu\rho^{ae}}(x,x') \nonumber \\
&&+gf^{abc}G_{\mu\rho}^{be}(x,x')\left(\partial^\mu (\eta^c_\nu\chi(x))-\partial_\nu(\eta^{\mu c}\chi(x))\right) \nonumber \\
&&+gf^{abc}\eta_\mu^b\chi(x)\left(\partial^\mu G_{\nu\rho}^{ce}(x,x')
-\partial_\nu G_{\mu\rho}^{ce}(x,x')\right) \nonumber \\
&&+gf^{abc}\partial^\mu \left(\eta^c_\nu\chi(x) G_{\mu\rho}^{be}(x,x')\right)
+gf^{abc}\partial^\mu\left(\eta_\mu^b\chi(x) G_{\nu\rho}^{ce}(x,x')\right) \\ \nonumber
&&+g^2f^{abc}f^{cdh}G_{\mu\rho}^{be}(x,x')\eta^{\mu d} \eta^h_\nu\chi^2(x) \\ \nonumber
&&+g^2f^{abc}f^{cdh}\eta^{b\mu}G_{\mu\rho}^{de}(x,x') \eta^h_\nu\chi^2(x) \\ \nonumber
&&+g^2f^{abc}f^{cdh}\eta^{b\mu}\eta^d_\mu G_{\nu\rho}^{he}(x,x')\chi^2(x)
= \delta_{ae}\eta_{\nu\rho}\delta^d(x-x').
\end{eqnarray}
In order to compute the propagator, we perform a gauge's choice. The most common is the Landau gauge ($\alpha=1$) that also grants that we are using exact formulas rather than asymptotic ones. So, we write as usual for this gauge
\begin{equation}
  G_{\mu\nu}^{ab}(x,x')=\delta_{ab}\left(g_{\mu\nu}-\frac{p_\mu p_\nu}{p^2}\right)\Delta(x,x')
\end{equation}
being $p_\mu$ the momentum
vector.
This yields for the above equation
\begin{equation}
   \partial^2\Delta(x,x')+3Ng^2\chi^2(x)\Delta(x,x')=\delta^d(x-x')
\end{equation}
that is the equation we were looking for. This equation coincides with that of the Green function of the scalar field obtained in \cite{Frasca:2013tma} in agreement with the mapping we derived in \cite{Frasca:2009yp} provided $\lambda\leftrightarrow Ng^2$, being $\lambda$ the corresponding coupling for the scalar field theory.


We now limit our analysis to the case $d=2+1$ and compute the exact Green function for this problem. The technique we follow is that outlined in Ref.\cite{Frasca:2013tma}. We just note that we have two independent solutions of the homogeneous equation
\begin{equation}
   \partial^2 y(x)+3Ng^2\chi^2(x)y(x)=0,
\end{equation}
or
\begin{equation}
   \partial^2 y(x)+3a^2\sqrt{2}(Ng^2)^2\operatorname{sn}^2\left(k\cdot x+\phi,-1\right)y(x)=0.
\end{equation}
One is
\begin{equation}
\label{eq:sol1}
   y_1(t)={\rm cn}(p\cdot x+\phi,-1){\rm dn}(p\cdot x+\phi,-1),
\end{equation}
with cn and dn elliptic Jacobi functions,
that holds provided
\begin{equation}
\label{eq:ds1}
    p^2=a^2\frac{N^2g^4}{\sqrt{2}}.
\end{equation}
The other one can be obtained by writing it as
\begin{equation}
   y_2(x)=y_1(x)\cdot w(x)
\end{equation}
with
\begin{equation}
   {\rm cn}(p\cdot x+\phi,-1){\rm dn}(p\cdot x+\phi,-1)\partial^2 w-4{\rm sn}^3(p\cdot x+\phi,-1)p\cdot\partial w=0.
\end{equation}
Now, we introduce a new variable $\bar x=p\cdot x+\phi$ and use the dispersion relation (\ref{eq:ds1}) to obtain
\begin{equation}
   {\rm cn}(\bar x,-1){\rm dn}(\bar x,-1)w''-4{\rm sn}^3(\bar x,-1)w'=0
\end{equation}
where the primes mean derivative with respect to $\bar x$. From eq.(\ref{eq:sol1}) we can obtain the solution in the rest reference frame $p_1=p_2=0$ and $p_0=aNg^2/2^\frac{1}{4}$. The corresponding Green function is
\begin{equation}
    G_R(t)=-\frac{1}{\mu_0 2^\frac{3}{4}}\theta(t){\rm cn}(\mu_0 t +\phi,-1){\rm dn}(\mu_0 t+\phi,-1)
\end{equation}
where we have set
\begin{equation}
   \mu_0=aNg^2/2^\frac{1}{4},
\end{equation}
%
that fixes the mass scale,
and $\theta(t)$ is the Heaviside function granting that the solution is different from 0 at $t>0$ and 0 for $t<0$ and provided that ${\rm cn}(\phi,-1)=0$. Similarly, one can define a backward propagating Green function as
\begin{equation}
    G_A(t)=\theta(-t){\rm cn}(-\mu_0 t +\phi,-1){\rm dn}(-\mu_0 t+\phi,-1).
\end{equation}
So, the propagator is
\begin{equation}
    G(t,0)=\delta^{d-1}(x)\left[G_A(t)+G_R(t)\right].
\end{equation}
When we turn to a Fourier transform, Fourier series of Jacobi functions are well-known \cite{NIST} giving
\begin{eqnarray}
    {\rm cn}(\mu_0 t +\phi,-1){\rm dn}(\mu_0 t+\phi,-1) &=& \frac{\pi^2}{K^2(-1)}\sum_{n=0}^\infty(-1)^n(2n+1)
   \frac{e^{-\left(n+\frac{1}{2}\right)\pi}}{1+e^{-(2n+1)\pi}}\times \nonumber \\
   \cos\left((2n+1)\frac{\pi}{2K(-1)}\mu_0 t+(2n+1)(4m+1)\frac{\pi}{2}\right)&&.
\end{eqnarray}
and so one arrives, back to the moving reference frame, at the result
\cite{Frasca:2013tma}
\begin{equation}
\label{eq:prop}
   G(p)=\sum_{n=0}^\infty\frac{B_n}{p^2-m_n^2+i\epsilon}
\end{equation}
with
\begin{equation}
    B_n=(2n+1)^2\frac{\pi^3}{4K^3(-1)}\frac{e^{-(n+\frac{1}{2})\pi}}{1+e^{-(2n+1)\pi}}.
\end{equation}
being $K(-1)$ the complete elliptic integral of the first kind and we get the ``mass spectrum''
\begin{equation}
\label{eq:ms}
   m_n=(2n+1)\frac{\pi}{2K(-1)}\mu_0.
\end{equation}
At this stage this has just a formal meaning. Moving to quantum field theory, we will prove that this is indeed the spectrum of the theory.
So, our final result for the Green function in $d=2+1$ is
\begin{equation}
    G_{\mu\nu}^{ab}(p)=\delta_{ab}\left(g_{\mu\nu}-\frac{p_\mu p_\nu}{p^2}\right)G(p).
\end{equation}
This result implies that the Yang-Mills theory shows up a mass gap also in this case. The corresponding spectrum can be used to fit with lattice data.

\section{Ghost sector \label{sec22}}

As shown in \cite{Frasca:2015yva} in four dimensions, starting from the exact solutions given in Sec.~\ref{sec1} for the 1-point function in the Dyson-Schwinger set of equations, the ghost propagator reduces just to the one of a free massless theory. This signals that the ghost sector decouples from the physical degrees of freedom. For the sake of completeness, we give here the corresponding Dyson-Schwinger equation that is
\begin{equation}
\partial^2 P^{am}_2(x-y)+gf^{abc}\partial^\mu(K^{bcm}_{3\mu}(0,x-y)+P^{bm}_2(x-y)G_{1\mu}^{c}(x)+P^{b}_1(x)K_{2\mu}^{cm}(x-y))=\delta_{am}\delta^3(x-y) 
\end{equation}
where we identify the 3-point and 2-point function $K^{bcm}_{3\mu}(0,x-y),\ K_{2\mu}^{cm}(x-y)$ for the ghost-gluon field propagation and the 1-point function $P^{b}_1(x)$ for the ghost field. With the given solutions for the 1-point function, this just boils down to the propagator for a free field.

\section{Quantum corrections \label{sec3}}

We want to see how quantum theory modifies the one- and two-point functions we obtained in the classical theory. This can be accomplished using the Dyson-Schwinger equations. We will stop the analysis to the two-point function as already discussed in our recent work \cite{Frasca:2015yva}. In that paper it is shown that the mass should be renormalized by
adding
the term
(given in $d$ dimensions)
\begin{equation}
  \delta\mu^2=3Ng^2\int\frac{d^dp}{(2\pi)^d}\sum_{n=0}^\infty\frac{B_n}{p^2-m^2_n+i\epsilon}
\end{equation}
where the dimensions of the coupling $g$ grant that of the squared mass. This integral can be exactly evaluated to give
\begin{equation}
   \delta\mu^2=\frac{3Ng^2}{(4\pi)^\frac{d}{2}}\Gamma\left(1-\frac{d}{2}\right)\sum_{n=0}^\infty B_n(m_n^2)^{\frac{d}{2}-1}.
\end{equation}
This correction diverges for $d=4$ while is finite for $d=3$. This should be expected for Yang-Mills theory in three dimensions \cite{Huber:2016tvc,Huber:2014tva}. It evaluates to
\begin{equation}
    \delta\mu^2=-\frac{3Ng^2}{4\pi}\sum_{n=0}^\infty B_n m_n\approx -a\frac{3N^2g^4}{2^\frac{9}{4}\pi}S_0
\end{equation}
with $S_0=2.046970223\ldots$ the result of the sum. This boils down to add a numerical constant to the arbitrary parameter $a$.
So, one can always redefine the factor $a$ in such a way to compensate the numerical factor obtained in this way and we will have for the spectrum
\begin{equation}
	m_n^2=(2n+1)^2\frac{\pi^2}{4K^2(-1)}a^2\frac{N^2g^4}{\sqrt{2}}.
\end{equation}

\section{Wilson loop and potential \label{sec4}}

In order to compute the potential in a pure Yang-Mills theory at the infrared fixed point, we have to evaluate
\begin{equation}
    \left\langle {\rm tr}{\cal P}e^{ig\oint_{\cal C} dx^\mu T^aA^a_\mu(x)}\right\rangle = \frac{\int [dA][d\bar c][dc] e^{-\frac{i}{4}\int d^3x{\rm Tr}F^2+iS_g[\bar c, c]}
    {\rm tr}{\cal P}e^{ig\oint_{\cal C} dx^\mu T^aA^a_\mu(x)}}{\int [dA][d\bar c][dc] e^{-\frac{i}{4}\int d^3x{\rm Tr}F^2+iS_g[\bar c, c]}}
\end{equation}
being $S_g[\bar c, c]$ the contribution of the ghost field, $T_a$ the anti-hermitian generators of the gauge group and ${\cal P}$ the path ordering operator. In our case, in the infrared limit, we have a trivial fixed point and the contribution coming from the Yang-Mills field is just a Gaussian one. This implies that our generating functional takes also a Gaussian form and
the Wilson loop has the simple form
\begin{equation}
     W[{\cal C}]\approx \exp\left[-T\frac{g^2}{2}C_2(R)\int\frac{d^2p}{(2\pi)^2}G({\bm p},0)e^{-i{\bm p}\cdot{\bm x}}\right]
\end{equation}
being $C_2(R)$ the quadratic Casimir operator that for SU(N) in the fundamental representation, $R=F$, is $C_2(F)=(N^2-1)/2N$. This yields
\begin{equation}
     W[{\cal C}]=\exp\left[-TV_{YM}(r)\right]
\end{equation}
being
\begin{equation}
     V_{YM}(r)=-\frac{g^2}{2}C_2(R)\int\frac{d^2p}{(2\pi)^2}G({\bm p},0)e^{-i{\bm p}\cdot{\bm x}}.
\end{equation}
%
We aim to recover the result given in \cite{Leigh:2005dg,Leigh:2006vg}. So, in our case we have to evaluate the integral \cite{Gonzalez:2011zc}
\begin{equation}
\label{eq:V1}
    V_{YM}(r)=-\frac{g^2}{4\pi}C_2(R)\int_0^\infty dp p G({\bm p},0)J_0(pr)
\end{equation}
being $J_0$ a Bessel function and with the propagator given in eq.~(\ref{eq:prop}). 
The integral can be computed exactly giving
\begin{equation}
   V_{YM}(r)=-\frac{g^2}{4\pi}C_2(R)\sum_{n=0}^\infty B_nK_0(m_nr)
\end{equation}
being $K_0$ a Bessel function. We recognize here the potential obtained, after some hypotheses, in \cite{Leigh:2005dg,Leigh:2006vg}. What is changing is the mass scale but this should be expected due to our approach that involves exact solutions to the classical equations of motion.
This potential grants that the three-dimensional theory is marginally confining \cite{Leigh:2005dg,Leigh:2006vg}.
%
The reason relies on the fact that $K_0(m_nr)\approx -\ln(m_nr/2)$ at smaller distances and decreases really slow with the distance making the Wilson loop not strictly proportional to the area. Anyhow, a logarithmic potential grants that the Gauss law is satisfied in 2+1 dimensions yielding colorless states.
%

We notice that
\begin{equation}
   V_{YM}(r)=-\sigma_{KKN}\frac{\pi}{2^\frac{5}{4}K(-1)}a\sum_{n=0}^\infty (2n+1)\frac{B_n}{m_n}K_0(m_nr)
\end{equation}
with $\sigma_{KKN}=\frac{N^2g^4}{8\pi}(1-1/N^2)$ the Karabali-Kim-Nair string tension.
%
Written in this way, we can compare it with the same result given in \cite{Leigh:2005dg,Leigh:2006vg}. We get a renormalized Karabali-Kim-Nair string tension given by
\begin{equation}
   \sigma^R_{KKN}=\sigma_{KKN}\frac{\pi}{2^\frac{5}{4}K(-1)}a.
\end{equation} 
The value of the arbitrary factor $a$ is irrelevant here as the square root of the string tension determines the spectrum through the ratio $m_n/\sqrt{\sigma^R_{KKN}}$ and this factor enters also into $m_n$ absorbing it.
Then, in the fundamental representation, 
we can take $a=1$ 
to compare with lattice data and one has
\begin{equation}
   \frac{\sqrt{\sigma}}{Ng^2} =\sqrt{1-\frac{1}{N^2}}\sqrt{\frac{1}{8\pi}}Z_\sigma^\frac{1}{2} = 0.2002189349-\frac{0.1001094674}{N^2}+O\left(\frac{1}{N^4}\right).
\end{equation}
where we have set $Z_\sigma=\frac{\pi}{2^\frac{5}{4}K(-1)}a$. This result agrees within an error of about 2\% with respect to lattice computations for the leading order in $1/N$ \cite{Athenodorou:2016ebg}.


\section{Glueball spectrum \label{sec5}}

The glueball spectrum for $0^{++}$ is easily obtained through the equation
\begin{equation}
   \frac{m_n}{\sqrt{\sigma}}=(2n+1)\frac{\pi}{2\cdot 2^\frac{1}{4}K(-1)}a\frac{Ng^2}{\sqrt{\sigma}}.
\end{equation}
%
Now, the factor $a$ entering into the string tension simplify with the one in the spectrum giving an overall value that, in some models in literature, was identified as a fudge factor but that here is fully justified by the exact solutions of the theory.
Indeed, the spectrum yields
\begin{equation}
	\frac{m_n}{\sqrt{\sigma}}=(2n+1)\cdot 5.032050686\ldots\cdot\sqrt{a}\frac{1}{\sqrt{1-\frac{1}{N^2}}}
\end{equation}
The agreement with the ground state of the theory is reached for $a=2/3$, assuming that the lattice data are affected by errors. In this case, the following table holds
\begin{table}[H]
\begin{center}
\begin{tabular}{|c|c|c|c|} \hline\hline
$N$      & Lattice    & Theoretical & Error \\ \hline
2        & 4.7367(55) & 4.744262871 & 0.16\% \\ \hline 
3        & 4.3683(73) & 4.357883714 & 0.2\% \\ \hline
4        & 4.242(9)   & 4.243397712 & 0.03\% \\ \hline
$\infty$ & 4.116(6)   & 4.108652166 & 0.18\% \\ \hline\hline
\end{tabular}
\caption{\label{tab:0++} Comparison for the ground state at varying $N$ and for $N\rightarrow\infty$
(lattice data are taken from \cite{Athenodorou:2016ebg}).
}
\end{center}
\end{table}
The agreement is strikingly good being well below 1\% error for the ground state at any $N$.
The factor emerging from the analysis of the ground state of the theory is in agreement with similar factors introduced in literature \cite{Isgur:1984bm,Johnson:2000qz} but now theoretically well founded.

\section{Conclusions \label{sec6}}

We have shown that, in the framework of our formalism,
marginal
confinement is achieved for QCD in three dimensions. We have found extensive agreement with lattice data and preceding theoretical works. We have also shown that numerical factors arbitrarily introduced in some models are completely justified by the set of classical solutions we have chosen to start with. The
exceptionally good
agreement between lattice data and theoretical predictions we achieved in the present case can serve as a justification {\sl a posteriori} for the choice of the solutions to start quantum field theory.


\end{document}